\documentstyle{article}
\begin{document}
\rightline{Jan 1994}
\rightline{McGill/94-9}
\vskip 2cm
\begin{center}
\begin{large}
A note on Koide's lepton mass relation
\vskip 2cm
\end{large}
R. Foot\footnote{email: Foot@hep.physics.mcgill.ca}\\
Department of Physics,\\
McGill University,\\
3600 University Street,\\
Montreal, Quebec, Canada.\\

\end{center}
\vskip 1cm
ABSTRACT
\vskip .5cm
\noindent
Koide has pointed out that the mass relation $m_e + m_{\mu} + m_{\tau}
= {2 \over 3}(\sqrt{m_e} + \sqrt{m_{\mu}} + \sqrt{m_{\tau}})^2 $
is consistent with the most recent measurements of the tau lepton mass.
We point out that this relation has a geometric interpretation.

\newpage

The origin of the fermion masses is perhaps the most pressing problem in
particle theory. The standard model can incorporate the fermion masses,
but it cannot provide any quantitative (or even qualitative) understanding.

Koide [1] has pointed out that the most recent measurements of the tau
lepton mass by the ARGUS, BES and CLEO
collaborations [2] ($1776.3 \pm 2.4 \pm 1.4$ MeV,
$1776.9 ^{+0.4}_{-0.5} \pm 0.2$ MeV,
and $1777.6 \pm 0.9 \pm 1.5$ MeV respectively)
are inagreement with the prediction 1776.97 MeV obtained from the lepton
mass formula
$$m_e + m_{\mu} + m_{\tau} = {2 \over 3} (\sqrt{m_e} + \sqrt{m_{\mu}}
+ \sqrt{m_{\tau}})^2. \eqno (1)$$
This seems a little surprising given the acuracy of the tau lepton
mass measurements. In fact, the above relation is a relation among the
physical masses. It does'nt seem possible for the same relation to hold
for the bare mass parameters because the electromagnetic corrections are
too big for the same relation to hold for the bare parameters. For this
reason, it seems difficult to understand the above relation from any
simple gauge model since it seems the electromagnetic corrections are
inevitable.

It could be that the mass formula eq.(1) is of no fundamental significance
since it is possible that its emperical success is just a
numerical coincidence.
Alternatively it could be that the forumla has something to do with
the real world. Since no-one knows the origin of the lepton masses this
is possible despite the fact that the formula cannot hold for the bare
masses. If anyone ever understands quantitatively the fermion masses  then
it will probably be achieved with guidence from experiment.
It will be interesting to see if the formula continues to be successful
when the tau mass value is further improved.

In this note, I wish to point out that Koide's mass relation has an
interpretation geometrically which makes it appear, in a sense, more
natural. I don't know if my way of looking at it is more useful or not.

Consider an ordinary three dimensional vector space. On this space
we put cartesian co-ordinates. Then the point
($\sqrt{m_e}, \sqrt{m_{\mu}}, \sqrt{m_{\tau}})$ and the origin define
a vector. Define $\theta$ as the angle between this vector and the
symmetric vector (the symmetric vector is the one which passes through
the origin and the point (1,1,1)). Then Koide's mass relation is equivalent
to the statement that $\theta = \pi/4$.
To see this, note that by definition,
$$\cos \theta = {(\sqrt{m_e}, \sqrt{m_{\mu}}, \sqrt{m_{\tau}}).(1,1,1) \over
|(\sqrt{m_e}, \sqrt{m_{\mu}}, \sqrt{m_{\tau}})||(1,1,1)|}, \eqno (2)$$
i.e.
$$\cos\theta = {\sqrt{m_e} + \sqrt{m_{\mu}} + \sqrt{m_{\tau}}
\over \sqrt{m_e + m_{\mu} + m_{\tau}} \sqrt{3}}. \eqno (3)$$
When $\cos\theta = 1/\sqrt{2}$ then eq.(3) reduces to Koide's formula
eq.(1). Theoretically the only constraint on $\theta$ (assuming
that at least one lepton has a non-zero mass) is
that $\cos \theta > 1/\sqrt{3}$ ( which means that $\theta < 54.7$ degrees).
Putting in the measured lepton mass values, then
$$\theta = 45.000 \pm 0.001. \eqno (3)$$
Note that if we replaced the $2/3$ factor in Koide's formula with
another factor then $\theta$ is nothing special (e.g.  if we
replaced the 2/3 with  3/4 or 1/2
then this corresponds with $\theta = 48.189...$ and $35.264...$ respectively).
If Koide's  formula has something to do with the real world, then
maybe this ``geometical formulation''  of the mass relation may
be useful inorder to derive the formula from some underlying theory.

\newpage
\vskip 1cm
\noindent
References

\vskip 1cm
\noindent
[1] Y. Koide, Mod. Phys. Lett. A8, 2071 (1993).
\vskip .4cm
\noindent
[2] H. Albrecht et al., ARGUS Collab., Phys. Lett. B292, 221 (1992);
J. Z. Bai et al., BES Collab., Phys. Rev. Lett. 69, 3021 (1992);
M. Daoudi et al., CLEO Collab., Talk given at XXVI Int. Conf. on High
Energy Physics, Dallas 1992.

\end{document}